\documentclass[prb,twocolumn,floatfix,notitlepage,superscriptaddress,longtable]{revtex4-2}
\usepackage{amsmath}
\usepackage{lipsum} 
\usepackage{verbatim}
\usepackage{epsfig}
\usepackage{subfigure}
\usepackage{graphicx}
\usepackage{amsfonts}
\usepackage[figuresright]{rotating}
\usepackage{amssymb}
\usepackage{amsmath}
\usepackage[normalem]{ulem}
\usepackage{psfrag}
\usepackage{esint} 
\usepackage{bm}
\usepackage[colorlinks,linkcolor=blue,anchorcolor=blue,citecolor=blue,urlcolor=blue]{hyperref}
\usepackage[version=4]{mhchem}
\usepackage[svgnames]{xcolor}

\def\be{\begin{equation}} \def\ee{\end{equation}}
\def\bea{\begin{eqnarray}} \def\eea{\end{eqnarray}}

\renewcommand{\vec}[1]{\mathbf{#1}}

\newcommand{\ket}[1]{| #1 \rangle}
\newcommand{\bra}[1]{\langle #1 |}

\def\bpm{\begin{pmatrix}} \def\epm{\end{pmatrix}}

\newcommand{\edited}[1]{{\color{Red}}}
\definecolor{Qicolor}{RGB}{3, 136, 252}

\makeatletter
\newcommand*{\balancecolsandclearpage}{%
  \close@column@grid
  \clearpage
}
\makeatother

\begin{document}
\title{Real-space topological invariant for time-quasiperiodic Majoranas}
\author{Zihao Qi}
\affiliation{Department of Physics, Cornell University, Ithaca, NY 14853, USA.}
\author{Ilyoun Na}
    \affiliation{Department of Physics, University of California, Berkeley, California 94720, USA}
    \affiliation{Materials Sciences Division, Lawrence Berkeley National Laboratory, Berkeley, California 94720, USA}
    \affiliation{Molecular Foundry, Lawrence Berkeley National Laboratory, Berkeley, California 94720, USA}
\author{Gil Refael}
\affiliation{Institute of Quantum Information and Matter and Department of Physics,California Institute of Technology, Pasadena, CA 91125, USA}
\author{Yang Peng}
\email[Corresponding author: ]{yang.peng@csun.edu}
\affiliation{Department of Physics and Astronomy, California State University, Northridge, Northridge, California 91330, USA}
\affiliation{Institute of Quantum Information and Matter and Department of Physics,California Institute of Technology, Pasadena, CA 91125, USA}
\date{\today}

\begin{abstract}
When subjected to quasiperiodic driving protocols, superconducting systems have been found to harbor robust time-quasiperiodic Majorana modes, extending the concept beyond static and Floquet systems. However, the presence of incommensurate driving frequencies results in dense energy spectra, rendering conventional methods of defining topological invariants based on band structure inadequate. In this work, we introduce a real-space topological invariant capable of identifying time-quasiperiodic Majoranas by leveraging the system's spectral localizer, which integrates information from both Hamiltonian and position operators. Drawing insights from non-Hermitian physics, we establish criteria for constructing the localizer and elucidate the robustness of this invariant in the presense of dense spectra. Our numerical simulations, focusing on a Kitaev chain driven by two incommensurate frequencies, validate the efficacy of our approach.
\end{abstract}

\maketitle

\section{Introduction}
The study of topological states under non-equilibrium conditions, particularly under a time-dependent potential, has been extensively explored over the past decade within the Floquet paradigm \cite{Oka2019,Rudner2020,Harper2020,Bao2022}, with the external drive being time-periodic. A notable development along this line of inquiry involves its extension to the time-quasiperiodic realm, in which quantum systems are subjected to external drives characterized by multiple mutually incommensurate frequencies \cite{Verdeny2016}. This extension allows for further control and manipulation of quantum systems and brings about a bunch of new topological phenomena, such as topological energy pumping due to the nontrivial topology in synthetic dimensions \cite{Martin2017,Peng2018,Crowley2019,Crowley2020,Nathan2021,Long2021,Qi2021,Boyers2020,Malz2021}. 

Moreover, time-quasiperiodic drives can also be used to generate nontrivial topology in physical dimensions with protected boundary modes. As first explored in Ref.~\cite{QPMajorana}, it was shown that a time-quasiperiodic Kitaev chain can host multiple types of Majorana boundary modes. In that work, the authors analyzed the quasienergy band structure using the enlarged Hamiltonian in the multi-frequency-extended space, which generalizes the Sambe space in the Floquet case \cite{sambe}. In particular, for the analysis a very small cutoff was chosen for the frequency-domain truncation, so that one is able to identify all the topological gaps opened at the intersection between bands dressed by multiple of driving frequencies, when periodic boundary condition (PBC) is assumed. Thus, one expects to have the Majorana edge modes inside these gaps when the system has an open boundary condition (OBC).  

As one increases the frequency cutoff, the spectrum of the enlarged Hamiltonian becomes denser and denser. Eventually, when one pushes the cutoff to infinity, one should expect a completely dense spectrum, in which for any eigenstate, there exists another one arbitrarily close to it in energy. Due to the absence of gaps, one should not expect to see any topological edge modes. 
Yet in the numerical calculations of Ref.~\cite{QPMajorana}, when the frequency cutoff was chosen large enough, the Majorana boundary modes did appear, thanks to the frequency-space localization, as explained by the authors. 

Despite the existence of the Majorana modes,  what was not addressed in that work is how to define a topological invariant that detects these boundary modes embedded in the dense spectrum. In the absence of gaps, certainly the Bloch-band-based topological invariants for the gapped band structures as in static and Floquet topological systems \cite{Yao2017} would not be applicable.  

In this work, we solve this problem by defining such a topological invariant inspired by a recently developed tool called \emph{spectral localizer} \cite{KTheory, Loring2017,Loring2020}, which has emerged as a versatile tool for probing real-space topology in a variety of materials with a gapless spectrum \cite{InvMetal, disorderedSemimetal, diffusiveMetal, trivialvTopoMetal, disorderedSemimetal,photonicGapless,topoPhotonicsGapless}, including disordered semimetals \cite{disorderedSemimetal}, as well as photonic structures \cite{topoPhotonicsGapless, photonicGapless}. 
Indeed, we are motivated by the fact that the spectral localizer can detect topological edge modes despite the presence of degenerate bulk bands, as appeared in the above mentioned systems. 

Given the above properties, we show that the spectral localizer is able to detect the topologial boundary modes in a time-quasiperiodic quantum system. We further present a physical interpretation of the invariant derived from the spectral localizer by making a connection to the physics of non-Hermitian Hamiltonians, which effectively describe dissipative systems \cite{Ashida2020}. Our work provides a concrete physical interpretation for the abstract spectral localizer within the context of non-Hermitian quantum systems. Particularly, we establish a criterion for selecting the free tuning parameter in defining the spectral localizer, thereby making this technique more practical. 

As an example, we shall consider the time-quasiperiodic Kitaev chain with particle-hole symmetry used in Ref.~\cite{QPMajorana}, and identify the local topological invariant.  Furthermore, we show the local invariant can be used to detect topological phase transitions associated with the emergence and disappearance of Majorana edge modes, despite of the dense spectrum in the large frequency truncation limit. 

The rest of this manuscript is organized as follows. In Sec.~\ref{sec:localizer}, we provide a brief overview of the spectral localizer. In Sec.~\ref{sec:qpsystem}, we discuss how the spectral localizer formalism can be applied to time-quasi-periodic systems and derive an appropriate local invariant. We then demonstrate it as the topological invariant for non-Hermitian matrices,  and provide criterion for selecting the free tuning parameter in defining the spectral localizer.
Following that, we numerically study a quasi-periodically driven Kitaev chain and demonstrate how spectral localizer can be used to detect topological phase transitions. We conclude with a discussion in Sec.~\ref{sec:discussions}.

\section{Review of Spectral Localizer \label{sec:localizer}}
In this section, we first briefly review the spectral localizer and its properties. From a mathematical perspective, the spectral localizer is a composite operator that combines a number of incompatible observables and determine whether they can be continued to commuting without breaking the system's symmetries. It is also shown to be closely related to a system's numerical $K$-theory~\cite{KTheory, Loring2017}. Practically, the spectral localizer allows one to probe a finite system's real-space topology, and has been shown to be more numerically efficient than other real-space approaches, such as computing the Bott index~\cite{disorderedSemimetal, BottIndex}. Given these properties, the spectral localizer has been primarily used to detect boundary-localized modes and probe non-trivial local topology in a variety of materials and systems~\cite{disorderedSemimetal, topoPhotonicsGapless, InvMetal, CrystallineTopo, topoPhotonics, photonicGapless, diffusiveMetal, trivialvTopoMetal, topoFineStructure, spectralLocalization, crystallineSymm, higherOrderTopo}.

For a system in $d$ dimensions, the Hamiltonian $H$ and its position operators $X_{j=1,\dots,d}$ are generally incompatible, $[H, X_j] \neq 0$. The spectral localizer is constructed by combining individual eigenvalue equations using a non-trivial Clifford representation~\cite{quadratic}. Namely, it is defined as~\cite{KTheory}:
\begin{align}
    &L_{x_1,...x_d, E}\left(X_1, X_2, ..., X_d, H\right) \nonumber \\
    &= \kappa \sum_{j=1}^d (X_j -x_j I) \otimes \Gamma_j + (H - EI) \otimes \Gamma_{d+1}.
    \label{eq:localizer}
\end{align}
Here $\kappa > 0$ is a scaling constant, and $I$ is the identity matrix with appropriate dimensions. The set of matrices $\{\Gamma_1, \Gamma_2, ..., \Gamma_{d+1}\}$ satisfies $\Gamma_j^\dagger = \Gamma_j$, $\Gamma_i \Gamma_j = -\Gamma_j \Gamma_i$, $\Gamma_j^2 = I$.  Unlike eigenvalue equations, however, here $\vec{x} = (x_1, x_2, ..., x_d)$ and $E$ are \textit{inputs} to the spectral localizer, and they dictate where in position-energy space the localizer is probing. Moreover, $\vec{x}$ and $E$ do not necessarily have to be eigenvalues of $(X_1, X_2, ..., X_d)$ and $H$. 

Depending on the system's symmetries, different properties of the spectral localizer can be used to determine if the set of matrices $\{X_1 - x_1 I, ..., X_d - x_d I, H - EI \}$ can be continued to commuting. If there is obstruction to that continuation, then the system exhibits non-trivial topology at $(\vec{x}, E)$. Probing whether the continuation is possible then allows us to 
define a local invariant that classifies a system's local topology~\cite{InvMetal, topoPhotonics, KTheory}. Similar to ones proposed in topological band theory~\cite{periodictable}, the local invariant will be one of the three types: matrix signature ($\mathbb{Z}$ invariant), sign of determinant ($\mathbb{Z}_2$), or sign of Pfaffian ($\mathbb{Z}_2$). In this work, we focus on 1D topological superconductors in Class D with a $\mathbb{Z}_2$ invariant, which is shown in later sections to be related to the the sign of the determinant of one block of the localizer.

%

\section{Spectral Localizer in time-quasiperiodic Systems \label{sec:qpsystem}}
\subsection{Frequency-Domain Representation}
Let us first review frequency-domain representation of operators. This representation allows us to work with a static Hamiltonian $K$ instead of the original, time-dependent Hamiltonian $H(t)$, and therefore to use the spectral localizer, which is defined for time-independent Hamiltonians.

Consider first a time-periodic (Floquet) system with period $T$, where the Hamiltonian $H(t)$ satisfies $H(t) = H(t+T)$. Denoting the system's physical Hilbert space as $\mathcal{H}$, we may equivalently formulate the problem by introducing an enlarged Hilbert space, $\mathcal{K} = \mathcal{H} \otimes L^2(S^1)$ (also known as the Sambe space~\cite{sambe}). Here $L^2(S^1)$ denotes the space of square integrable $T$-periodic functions.

In the enlarged Hilbert space $\mathcal{K}$, a time-periodic state $\ket{\psi(t)}$ can be represented as:
\begin{equation}
    \ket{\psi(t)} = \sum_n \ket{\psi_n} e^{-in\omega t} \rightarrow \sum_n \ket{\psi_n} \otimes \ket{n},
\end{equation}
where $\ket{n}$ is a state in the Fourier space of $L^2(S^1)$. Intuitively, the time-periodic problem is mapped to a 1D synthetic lattice, where the index $n$ denotes the position on that lattice. Similarly, the Hamiltonian $H(t)$ and time derivative operator $-i \partial_t$ can be written as, respectively, $\sum_n h_n \otimes \ket{n}$ and $-\omega\mathbb{I}_{\mathcal{H}} \otimes \hat{N}$, where $\mathbb{I}_{\mathcal{H}}$ is the identity matrix on the physical Hilbert space, $\hat{N}:=\sum_n n \ket{n}\bra{n}$ is the position operator on the frequency lattice, and $\omega = 2\pi/T$ is the frequency. Therefore, in the enlarged space, the enlarged Hamiltonian is represented by the following static matrix $K$:
\begin{equation}
    H(t)-i\partial_t \rightarrow K := - \omega \mathbb{I}_{\mathcal{H}} \otimes \hat{N} + \sum_{n} h_n \otimes \sigma_n ,
    \label{eq:KFloquet}
\end{equation}
where we have also introduced $\sigma_n := \sum_m \ket{n+m}\bra{m}$ as the operator that shifts all sites on the Floquet lattice by distance $n$, namely $\sigma_n \ket{m} = \ket{n+m}$. 

As an example, consider a harmonically driven system, $H(t) = h_0 + 2 h_1 \cos(\omega t)$. The non-trivial harmonics are $h_0$ and $h_1 = h_{-1}^\dagger$. In this representation, the enlarged Hamiltonian is written explicitly as:
\begin{equation}
    K = \begin{pmatrix}
    \ddots \\
   &  h_0 - \omega & h_1 & 0 \\
   &  h_1^\dagger & h_0 & h_1 \\
   & 0 & h_1^\dagger & h_0 + \omega \\
    & & & & \ddots
    \end{pmatrix}.
\end{equation}
The quasi-energies can then be obtained by diagonalizing $K$. Note that the structure of $K$ immediately implies that the quasi-energies are only defined modulo $\omega$ in the Floquet system. In practice, we have to truncate $K$ to a finite number of sectors, and we will denote the cutoff as $M$.

The above discussion on Floquet systems generalizes easily to time-quasiperiodic systems, where $H(t)$ depends on $s$ mutually incommensurate frequencies. In this case, the emergent synthetic lattice $\ket{\boldsymbol{n}}$ will be $s$-dimensional, and the static Hamiltonian $K$ will take a form similar to that in Eq.~(\ref{eq:KFloquet}), with its sectors being:
\begin{equation}
K_{\boldsymbol{n},\boldsymbol{m}}=H_{\boldsymbol{n}-\boldsymbol{m}}-\delta_{\boldsymbol{n},\boldsymbol{m}}\boldsymbol{n}\cdot\boldsymbol{\omega},
\label{eq:qpsectors}
\end{equation}
where $\boldsymbol{n}, \boldsymbol{m} \in \mathbb{Z}^s$, and $\boldsymbol{\omega} = (\omega_1, \omega_2, ..., \omega_s)$ the vector of frequencies.

As a concrete example, for a system driven by two mutually irrational frequencies $\omega_1$ and $\omega_2$, the static Hamiltonian $K$ is:
\begin{align}
    K = -\omega_1 \, \mathbb{I}_{\mathcal{H}} \otimes \hat{N}^{(1)} \otimes \mathbb{I}^{(2)} -\omega_2 \, \mathbb{I}_{\mathcal{H}} \otimes \mathbb{I}^{(1)} \otimes \hat{N}^{(2)} \nonumber
    \\ + \sum_n h_n^{(1)} \otimes \sigma_n^{(1)} \otimes \mathbb{I}^{(2)} + \sum_n h_n^{(2)} \otimes \mathbb{I}^{(1)} \otimes \sigma_n^{(2)},
\end{align}
where the superscript denotes operators from the corresponding drive. Again from the first two terms, we see that quasi-energies in this system is only defined modulo $n_1 \omega_1 + n_2 \omega_2$, where $n_1, n_2 \in \mathbb{Z}$.

\subsection{Real-Space Topological Invariant}
Similar to Majoranas in static systems, in 1D time-quasiperiodic Majoranas are also protected by particle-hole symmetry (Class D).
In this section, we will identify the local invariant for such systems. 

With only one spatial dimension, we have a single position operator $X$, which acts trivially on the frequency lattice spanned by $\ket{\boldsymbol{n}}$. We construct the spectral localizer composing of operators $K$ and $X$
\begin{align}
    &L_{x, \epsilon}(X, K) = \kappa (X-xI) \otimes \sigma_x + (K-\epsilon I) \otimes \sigma_y \nonumber \\
    &= \begin{pmatrix}
    0 & \kappa (X-xI) - i (K-\epsilon I) \\
    \kappa (X-xI) + i (K-\epsilon I) & 0
    \end{pmatrix}. 
    \label{eq:XKlocalizer}
\end{align}
Here $x$ and $\epsilon$ are parameters with dimensions of position and energy, $\sigma_{x, y}$ are the Pauli $x$ and $y$ matrices satisfying the Clifford algebra, and $\kappa > 0$ is a scaling constant. It was stated in previous works that while the allowed values of $\kappa$ spans a wide range, it should be chosen below some critical value $\kappa_c$~\cite{Loring2017, Loring2020}. However, the meaning of $\kappa_c$ was not clearly stated, and the appropriate values of $\kappa$ were chosen only empirically~\cite{wong2024classifying}. In Sec.~\ref{sec:nonhermitian}, we shall present the meaning for this $\kappa_c$ in our system.

\textcolor{black}{In Class D, there is a particle-hole symmetry realized by a unitary matrix $V_c$ with $V_cV^{{\dagger}}_c=1$
that transforms the enlarged Hamiltonian as $V_c (K-\bar{\epsilon}I)^* V_c^{-1} = -(K - \bar{\epsilon}I)$.} Here $\bar{\epsilon}$ can be any particle-hole symmetric quasi-energies of the form $\bar{\epsilon} = \boldsymbol{l}\cdot\boldsymbol{\omega}$ with $\boldsymbol{l}$ a $s$-dimensional vector consisted of integers and half integers. Given multiple inequivalent particle-hole symmetric quasi-energies, multiple types of Majorana can be obtained \cite{QPMajorana}, and thus we need to detect Majoranas at different quasi-energies.

Because of the particle-hole symmetry, the spectral localizer is also particle-hole symmetric at $\epsilon=\bar{\epsilon}$
\begin{equation}
    (V_c \otimes \sigma_z) \, L^{*}_{x,\bar{\epsilon}} \, (V_c \otimes \sigma_z)^{-1} = -L_{x,\bar{\epsilon}}.
\end{equation}
Moreover, $L_{x,\epsilon}$ has an additional chiral symmetry introduced by the particular Clifford representation, 
\begin{equation}
    (I \otimes \sigma_z) \, L_{x,\epsilon} \, (I \otimes \sigma_z)^{-1} = -L_{x,\epsilon}.
\end{equation}

Given these symmetries, for each pair $(x, \bar{\epsilon})$, the spectral localizer is a Hermitian matrix describing an effective 0D system in Class BDI, and is classified by a $\mathbb{Z}_2$ invariant~\cite{nonlinearTopoMaterial}. To see this, we use the fact that 
via a basis transformation, $K-\bar{\epsilon}I$ can always be brought into a pure imaginary form whereas $X-xI$ is purely real. 
In this basis, we can define the topological invariant for $L_{x,\bar{\epsilon}}$
\begin{equation}
    C_{x,\bar{\epsilon}} = \text{sign}\left(\text{det}\left(\kappa(X-xI) + i(K-\bar{\epsilon}I)\right)\right).
    \label{eq:invariant}
\end{equation}
Note that the matrix $\kappa(X-xI) + i(K-\bar{\epsilon}I)$ is real, which guarantees the determinant to be real. This invariant is in agreement with, and a generalization of, its counterpart in static, particle-hole symmetric systems, in which the invariant is only defined at $\bar{\epsilon} = 0$~\cite{topoPhotonics, KTheory}. 

\textcolor{black}{To detect topologically protected boundary modes, such as the Majorana modes in the time-quasiperiodic topological superconductor introduced above, we need to examine 
the invariant $C_{x,\bar{\epsilon}}$ at fixed $\bar{\epsilon}$, which is determined by the type of time-quasiperiodic Majoranas we are interested in, and vary the position argument $x$. For small $\kappa$ and $x$ inside the bulk, 
$C_{x,\bar{\epsilon}}$ will take the value of the bulk topological invariant obtained from the topological band theory in a periodic system \cite{topoPhotonics, KTheory}. When $x$ is near the boundary, $C_{x,\bar{\epsilon}}=1$ is trivial. Thus, the real space signature
for the existence of a Majorana at quasienergy $\bar{\epsilon}$ is the jump from $1$ to $-1$ for $C_{x,\bar{\epsilon}}$ as $x$ is varied from the boundary into the bulk.}

Next, we introduce this basis transformation explicitly. The key insight is to find a basis in which $V_c$ is an identity operator
so that $(K - \bar{\epsilon}I) =-(K-\bar{\epsilon}I)^*$ is purely imaginary (and antisymmetric). 
To this end, we define the transformation matrix 
\begin{equation}
W=\frac{1}{\sqrt{2}}\left(\begin{array}{cc}
\mathbb{I}_{x} & i\mathbb{I}_{x}\\
\mathbb{I}_{x} & -i\mathbb{I}_{x}
\end{array}\right),
\end{equation}
where $\mathbb{I}_{x}$ is the identity matrix in real space. For particle-hole symmetry centered at quasi-energy $\bar{\epsilon} =\boldsymbol{l}\cdot\boldsymbol{\omega}$, we can further define the following matrix in frequency domain:
\begin{equation}
\rho_{\boldsymbol{n,m}}= \frac{1}{\sqrt{2}}\begin{cases}
1 & \boldsymbol{n=}\boldsymbol{m}<\boldsymbol{l}\\
-i & \boldsymbol{n}=\boldsymbol{m}>\boldsymbol{l}\\
\sqrt{2} & \boldsymbol{n=m=l}\\
i & \boldsymbol{n}=-\boldsymbol{m}<\boldsymbol{l}\\
1 & \boldsymbol{n}=-\boldsymbol{m}>\boldsymbol{l}
\end{cases}.
\end{equation}
where vectors are ranked by comparing their indices, starting from the first one. One can then check that the following matrix after basis transformation
\begin{equation}
(W \otimes \rho)^{\dagger}K(W \otimes \rho) - \bar{\epsilon} I
\end{equation}
is purely imaginary. On the other hand, since the transformation matrix $W \otimes \rho$ acts trivially on the spatial dimension,  $X - xI$ is unchanged under the transformation, i.e., it remains real and diagonal.

\subsection{Interpretation from non-Hermitian Physics \label{sec:nonhermitian}}
The invariant of the spectral localizer $L_{x,\bar{\epsilon}}$ defined in Eq.~(\ref{eq:invariant}) involves the non-Hermitian Parity-Time (PT)-symmetric real matrix $M_x(\kappa) = \kappa(X-xI) + i(K-\bar{\epsilon}I)$, where the PT symmetry is simply the complex conjugation. 
It is known that there are two types of eigenvalues of a PT-symmetric matrix: the PT-preserving ones (real) and the PT-breaking ones (complex)~\cite{yang2023homotopy}.
Note that the PT-breaking eigenvalues must appear as complex conjugate pairs $E,E^* \in \mathbb {C}$. 

Now, let us look at the matrix $M_x(\kappa)$ in our problem. At $\kappa=0$, $M_x(0) = i(K - \bar{\epsilon}I)$ is simply the enlarged Hamiltonian multiplied by the imaginary unit. Because of the particle-hole symmetry around $\bar{\epsilon}$, the eigenvalues of $M_x(0)$ are on the imaginary axis and appear in conjugation pairs as $ \pm i \epsilon_n$ corresponding to non-Majorana modes, and zero corresponding to possible Majoranas. Hence, the topologically trivial bulk states can be regarded as the PT-breaking states, whereas the Majoranas are the PT-preserving states. 

As $\kappa$ increases from zero, all PT-breaking eigenvalues must move in pairs with the same real part, $\pm i\epsilon_n \to \Gamma_n \pm i\epsilon'_n$. Note that the invariant in Eq.~(\ref{eq:invariant}) can be written as the sign of the product of all eigenvalues of $M_x(\kappa)$. 
Since the PT-breaking eigenvalues appear as complex conjugation pairs, their product is always real and positive. We thus conclude $C_{x,\bar{\epsilon}}=1$ is trivial if there are no Majoranas i.e. no PT-preserving states. 

However, if at $\kappa=0$ the system has two Majoranas at the left and right boundary, their eigenvalues can move independently along the real axis $(0, 0) \to (\delta_{L}, \delta_R)$ as $\kappa$ is increased~\cite{yang2023homotopy}. Since the product of all other eigenvalues is real and positive, the invariant can be written as
\begin{equation}
C_{x,\bar{\epsilon}} = \text{sign}(\delta_L \delta_R).
\end{equation}
For small $\kappa$, based on perturbation theory we have $\delta_{L,R} \simeq \kappa \bra{\psi_{L,R}}(X-xI)\ket{\psi_{L,R}}$, where $\ket{\psi_{L, R}}$ denotes the Majorana mode at the left and right boundary respectively. If $x$ is chosen near either boundary, the matrix $X - xI$ is either positive or negative semidefinite, which produces the same sign for $\delta_L$ and $\delta_R$ and gives $C_{x,\bar{\epsilon}}=1$. However, if $x$ is chosen deep inside the bulk, $\bra{\psi_L}X-xI\ket{\psi_L}$ is negative, while $\bra{\psi_R}X-xI\ket{\psi_R}$ is positive, which gives opposite signs for $\delta_L$ and $\delta_R$, and leads to $C_{x,\bar{\epsilon}}=-1$.

The connection to non-Hermitian systems also explains why the hyperparameter $\kappa$ cannot be too large when constructing the spectral localizer \cite{topoPhotonics,KTheory}. This can be understood from the spectrum of the matrix $M_x(\kappa)$. As $\kappa$ is increased, all our above statements remain true, until an exceptional point is reached when $\kappa=\kappa_c$, where the matrix $M_x(\kappa_c)$ becomes non-diagonalizable through the coalescence of pairs of eigenvalues and eigenvectors \cite{berry2004physics,heiss2012physics}.
For $\kappa>\kappa_c$, $M_x(\kappa)$ no longer hosts PT-preserving modes, indicating that it is topologically different from $M_x(0) \equiv iK$, which has Majoranas as PT-preserving modes. 
Thus, the criterion for selecting $\kappa$ is that no exceptional point should be met as $\kappa$ is increased from zero.

\subsection{Effects due to Dense Spectrum \label{sec:dense}}
One complication in time-quasiperiodic systems is the dense energy spectrum when one extends the frequency truncation limit to infinity.
This can lead to a vanishing ``\emph{localizer gap}" \cite{InvMetal}, which is defined as the smallest eigenvalue (in modulus) of $L_{x,\bar{\epsilon}}$ at each $(x,\bar{\epsilon})$. For our system, the localizer gap at each $x$ is equal to the smallest singular value of matrix $M_x(\kappa)$, namely $\min[\sigma_s(M_x(\kappa))]$, where $\sigma_s(\cdot)$ denotes the singular value spectrum of a matrix. 

To analyze the localizer gap of $M_x(\kappa)$, it is useful to consider the eigenvalues of $M_x(\kappa)$, which is different from its singular values. If we denote the smallest eigenvalue of $M_x(\kappa)$ in absolute value as $\epsilon_{\text{min}} \in \mathbb{C}$,
then we have in general $\min[\sigma_s(M_x(\kappa))]\leq |\epsilon_{\text{min}}|$. This means whenever $\epsilon_{\text{min}} = 0$, the localizer gap vanishes. Conversely, vanishing localizer gap implies $0\in \sigma_s(M_x(\kappa))$ and $\det(M_x(\kappa)) = 0$, thereby forcing $\epsilon_{\text{min}} = 0$. $\epsilon_{\text{min}}$ can therefore be regarded as an alternative to the localizer gap, given that the localizer gap closes if and only if $\epsilon_{\text{min}}=0$, where the invariant $C_{x,\bar{\epsilon}}$ is no longer well defined.

One advantage of analyzing the eigenvalues of $M_x(\kappa)$ is that they are related at different spatial locations $x$, because 
$M_x(\kappa) = M_{x=0}(\kappa) -\kappa xI$. In other words, the eigenvalues of $M_x(\kappa)$ at generic $x$ can be obtained from the ones at $x=0$ together with an additional shift $-\kappa x$. Without loss of generality, let us choose the coordinate system such that the left boundary is at the origin $x=0$, and sites along the chain are located at integer coordinates $x=0,1,\dots,N-1$. This means the position operator $X$ has eigenvalues $0,1,\dots,N-1$ and is thus positive semidefinite. We can denote eigenvalues of $M_{x=0}(\kappa)$
as $\Gamma_n \pm i(\epsilon_n +  \boldsymbol{m}\cdot\boldsymbol{\omega})$ and $\delta_{L,R} + i\boldsymbol{m}\cdot\boldsymbol{\omega}$, with
$\Gamma_n, \delta_{L,R}\geq 0$,
corresponding to PT-breaking states and PT-preserving states, respectively. Here the term $\boldsymbol{m}\cdot\boldsymbol{\omega}$ takes into account the time-quasiperiodicity represented in frequency domin~\cite{QPMajorana}. 

In the limit of large frequency truncation limit , $\boldsymbol{m}\cdot\boldsymbol{\omega}$ can approach any real value by choosing sufficiently large (in magnitude) integers in the integer-valued vector $\boldsymbol{m}$. 
Therefore, the imaginary parts of eigenvalues of $M_{x=0}(\kappa)$, namely $\epsilon_n+\boldsymbol{m}\cdot\boldsymbol{\omega}$ and $\boldsymbol{m}\cdot\boldsymbol{\omega}$, are dense in $\mathbb{R}$ (can be made arbitrarily close to any real number).
Particularly, these imaginary parts can approach zero. This implies
\begin{equation}
   \epsilon_{\text{min}} =  \min_{n;j=L,R}(|\Gamma_n - \kappa x|, |\delta_j - \kappa x|).
   \label{eq:epsilon_min}
\end{equation}
As $x$ increases from $0$, $\kappa x$ may become equal to a particular $\Gamma_n$ or $\delta_j$, and hence lead to $\epsilon_{\text{min}} = 0$ and vanishing localizer gap. Since the number of distinct real part of eigenvalues $\Gamma_n$ and $\delta_j$ is linear in $L$ (in contrast to the dense imaginary part), there is only a finite number of spatial locations $x > 0$ where the localizer gap and $\epsilon_{\text{min}}$ vanish. 

The above analysis shows that despite the dense spectrum of $K$, the invariant $C_{x,\bar{\epsilon}}$ is undefined only at a finite number of points (a measure zero set). Moreover, to calculate the $\epsilon_{\text{min}}$, we only need to get the distinct real part of the eigenvalues of $M(\kappa)$. Numerically, as shown in Appendix~\ref{app:MReal}, the real part easily converges with a small frequency-domain truncation.

The next question is: can we still use $C_{x,\bar{\epsilon}}$ to detect Majoranas by sweeping $x$ when there is a finite number of $x$ where $C_{x,\bar{\epsilon}}$ is not well-defined? To answer this question, we first note that vanishing localizer gap/$\epsilon_{\text{min}}$ does not necessarily imply a change in $C_{x,\bar{\epsilon}}$. 
In particular, the invariant does not change if the gap closing is due to a pair of PT-breaking eigenvalues going through zero,
namely $\Gamma_n = \kappa x$ for some $n$. This is because the product of this pair of eigenvalues remains positive, before and after closing. The invariant $C_{x,\bar{\epsilon}}$ only changes if a PT-\textit{preserving} eigenvalue goes through zero, namely $\delta_j = \kappa x$ for $j=L$ or $R$. Hence, in spite of many points of localizer gap closing, the transition between $1$ and $-1$ in $C_{x,\bar{\epsilon}}$ only happens when the gap closing is due to $\delta_{L}$ or $\delta_{R}$ crossing $\kappa x$. The usage of the spectral localizer is therefore still valid.

\section{Numerical Example: Driven Kitaev Chain \label{sec:numerical}}
\subsection{Quasi-Periodic Majoranas \label{sec:qpmajorana}}
To illustrate the ideas introduced previously, as an example we shall take the quasiperiodically driven Kitaev chain~\cite{QPMajorana}, which is known to host multiple types of time-quasiperiodic Majoranas.

The Hamiltonian for this driven Kitaev chain has the following form
\begin{equation}
    H(t) = H_K + \sum_i V(\omega_i t; \Delta_i),
    \label{eq:drivenchain}
\end{equation}
where 
\begin{equation}
    H_K = -\mu \sum_{j=1}^N c_j^\dagger c_j - \sum_{j=1}^{N-1} [(J c_j^\dagger c_{j+1} + i\Delta c_j c_{j+1}) + \text{h.c.}]
    \label{eq:staticchain}
\end{equation}
is the Hamiltonian for a static Kitaev chain~\cite{Kitaev2001}, where
$c_j (c_j^\dagger)$ is the annihilation (creation) operator on site $j$, $J$ is the nearest-neighbor hopping amplitude, $\mu$ the onsite potential, and $\Delta$
the static pairing potential.
The term
\begin{equation}
    V(\omega_i t; \Delta_i) = -i \Delta_i \sum_{j=1}^{N-1} (e^{-i\omega t} c_j c_{j+1} - e^{i \omega t} c_{j+1}^\dagger c_j^\dagger)
    \label{eq:drive}
\end{equation}
describes a time-periodic pairing term at frequency $\omega_i$, with amplitude $\Delta_i$.

\textcolor{black}{In Appendix~\ref{app:kitaev}, we review the condition for the existence of different time-quasiperiodic Majoranas in this model. It turns out that
by including two independent dynamic pairing potential at two frequencies $\omega_1$ and $\omega_2$, Majorana modes at quasi-energy $\omega_1/2$, $\omega_2/2$, together with the zero mode, can be engineered. In terms of system's parameters,  the $\omega_1/2$ and $\omega_2/2$ Majoranas disappear when $\omega_1$, $\omega_2$, exceed the bandwidth $2J + \mu$, respectively. On the other hand, the zero mode should always survive, as long as $|\mu| < 2|J|$, independent of $\omega_1$ and $\omega_2$. Moreover, the $(\omega_1+\omega_2)/2$-Majorana can be engineered if we include the dynamic pairing potential $V((\omega_1 - \omega_2)t; \Delta_{12})$ in addition to the other two dynamic pairing terms. It requires $|\omega_1 -\omega_2|< (2J+\mu)$, and $\omega_{1,2} \gtrsim (2J+\mu)$.}

\subsection{Numerical Results}
We numerically study the quasi-periodically driven Kitaev chain described by Eq.~(\ref{eq:drivenchain}), with two dynamic pairing potential \textcolor{black}{at frequencies $\omega_1$ and $\omega_2$}, using the spectral localizer $L_{x,\bar{\epsilon}}$ as defined in Eq.~(\ref{eq:XKlocalizer}) and the local invariant $C_{x,\bar{\epsilon}}$ defined in Eq.~(\ref{eq:invariant}).

\begin{figure}
    \includegraphics[width = \linewidth]{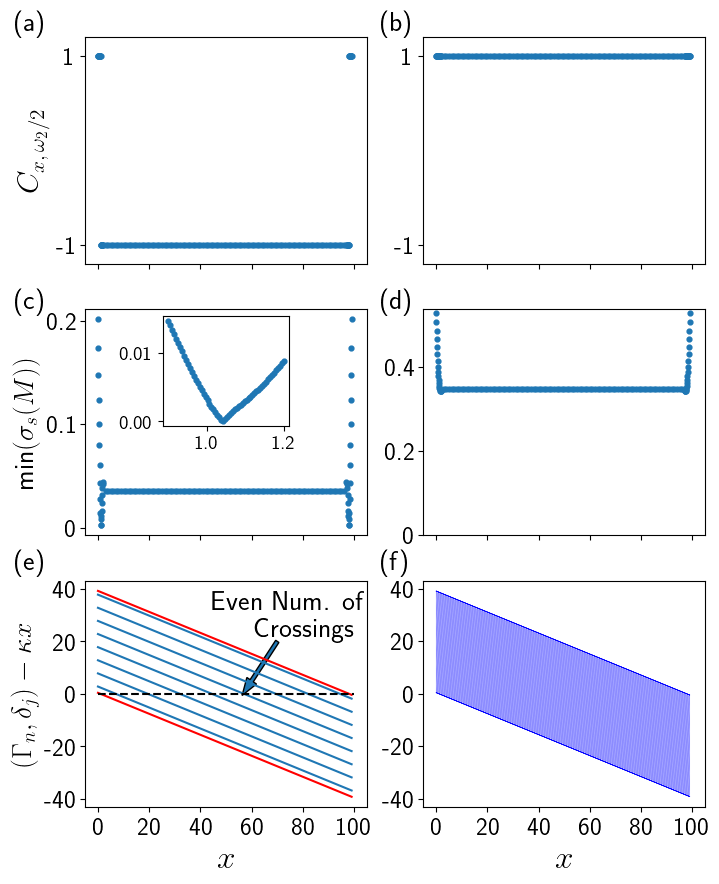}
\caption{(a, b) Local invariant $C_{x, \bar{\epsilon}}$ along the chain in the presence and absence of Majorana modes. The bulk invariant is non-trivial when topological edge modes are present. (c, d) localizer gap $\min(\sigma_s(M))$ in the presence and absence of Majoranas. Inset of (c): localizer gap zoomed-in around the system boundary. The location of localizer gap closing in (c) coincides with that of invariant change in (a). (e): Distinct real part of the eigenvalues of $M(\kappa)$. The Majorana modes ($\delta_j - \kappa x$) and a few bulk states ($\Gamma_n - \kappa x$) are shown in red and blue, respectively. The black dashed line is at $y = 0.$ Intersections with it indicates vanishing $(\Gamma_n, \delta_j) - \kappa x$. \textcolor{black}{Only the Majorana crossings, shown in red, correspond to invariant changes. Bulk states crossings, shown in blue, occur an even number of times and do not change the invariant.} (f): Real part of eigenvalues of $M(\kappa)$ in an energy window $(-\omega_2/2, \omega_2/2)$. While the imaginary part of the spectrum is dense, the real part is not.  The parameters used across all panels are: $J/\omega_1 = 0.5,\mu/\omega_1 = 1, \Delta/\omega_1 = 0.05, \Delta'/\omega_1 = 0.05, \kappa = 0.4, N = 100.$ The frequency ratio is $\omega_2/\omega_1 = (\sqrt{5}+1)/2$ (a, c, e, f), $\omega_2/\omega_1 = 5(\sqrt{5}+1)/2$ (b, d). The frequency domain is truncated to a $9 \times 8$ lattice.}
    \label{fig:main}
\end{figure}

First, we show the behavior of the local invariant $C_{x, \bar{\epsilon}}$ and the localizer gap, $\min(\sigma_s(M))$, in the presence and absence of Majorana modes. For concreteness, we have focused on the Majorana mode at quasi-energy $\omega_2/2$ by setting $\bar{\epsilon} = \omega_2/2$. As shown in Fig.~\ref{fig:main}(a, b), the local invariant becomes non-trivial in the bulk when the system hosts Majorana edge modes. On the other hand, when Majoranas are absent (specifically, for $\omega_2$ exceeding the bandwidth $2J+\mu$; see Appendix ~\ref{app:kitaev}), the invariant remains trivial along the chain. This correspondence between the existence (absence) of topological boundary modes and nontrivial (trivial) bulk invariants has also been observed in other systems using the spectral localizer approach, such as anomalous Floquet topological insulators~\cite{higherOrderTopo} and nonlinear topological materials~\cite{nonlinearTopoMaterial}.

The localizer gap behaves similarly, as shown in Fig.~\ref{fig:main} (c, d). In the presence of Majoranas, the localizer gap indeed closes near the boundary, while it remains open when the system is topologically trivial. The location of the gap closing coincides with that of invariant change shown in Fig.~\ref{fig:main}(a). Importantly, we would like to note that the appearance of a constant localizer gap in the bulk is merely an artifact of finite sampling of $x$: namely, we have sampled $x$ densely around the boundary and chosen $x$ on site in the bulk to single out the effects of Majorana edge modes. If we were to sample the bulk as densely, we would have expected many closings due to topologically trivial states. However, such crossings always occur in pair and therefore do not affect the local invariant, as discussed in Sec.~\ref{sec:dense}. 

To further support our analysis in Sec.~\ref{sec:nonhermitian} and \ref{sec:dense}, we have also analyzed the spectrum of $M_x(\kappa)$. In Fig.~\ref{fig:main}(e), we show the real part of a few eigenvalues of $M_x(\kappa)$ along the chain, i.e., $(\Gamma_n, \delta_j) - \kappa x$ for both $\delta_L, \delta_R$ and a few bulk states $\Gamma_n$. Compared to the local invariant shown in Fig.~\ref{fig:main}(a), it is clear that only the Majorana crossings at $\delta_j -\kappa x = 0$ change the local invariant. On the other hand, while there is an extensive number of crossings at $\Gamma_n - \kappa x = 0$ due to bulk states, the local invariant is not affected, as such crossings always occur in pairs. In Fig.~\ref{fig:main}(f) we show $(\Gamma_n, \delta_j) - \kappa x$ for \textit{all} $\Gamma_n$ and $\delta_j$ in an energy window $(-\omega_2/2, \omega_2/2)$. In the limit of large frequency-domain truncation, while the imaginary part of $M$'s spectrum becomes dense, the real part of the spectrum remains unchanged. Therefore, despite the dense spectrum, $\epsilon_{\text{min}}$ still serves as a useful alternative to the localizer gap. For the behavior of the localizer gap and $\epsilon_{\text{min}}$ in the limit of an infinite frequency lattice, see Appendix.~\ref{app:MReal}.

\begin{figure}
    \centering
\includegraphics[width=0.98\linewidth]{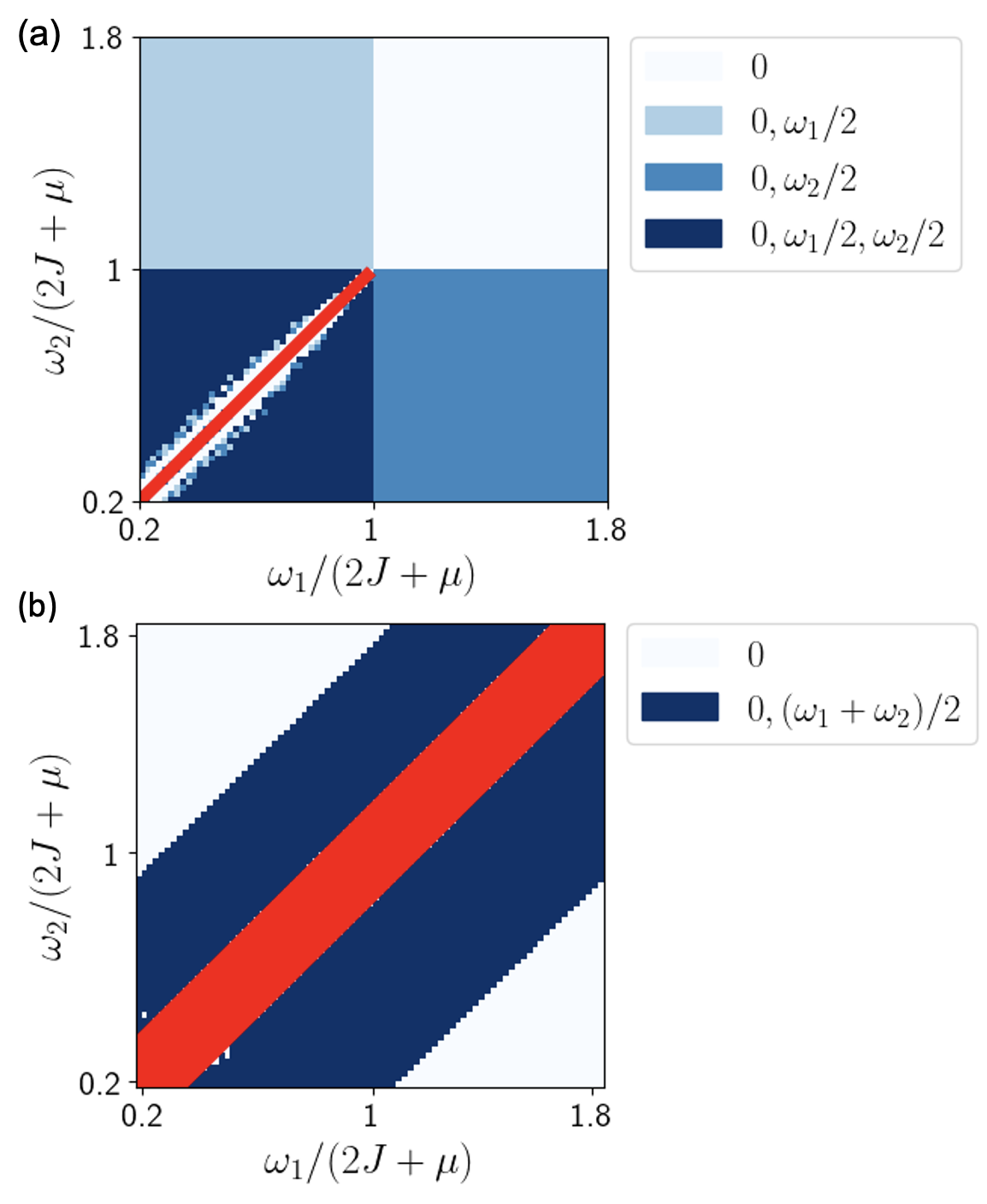}
    \caption{\textcolor{black}{Phase diagrams of co-existing Majorana modes from probing the bulk local invariant. (a) Phase diagram consisting of  three Majorana modes with quasi-energies $\bar{\epsilon} = 0, \omega_1/2, \omega_2/2$. (b) Same as (a), but with two modes at $\bar{\epsilon} = 0, (\omega_1 + \omega_2)/2$. In both panels, the red line indicates that the parameter regime  $\omega_1 \simeq \omega_2$ is not considered. The parameters used for both figures are $\mu/J = 1, \Delta/J = 0.1, \Delta_{1}/J = \Delta_2/J = 0.05, \kappa = 0.001, N = 60$. In 2(b), there is an extra pairing potential $\Delta_{12}/J = 0.05.$}}
    \label{fig:inv_freq}
\end{figure}

Now that we have established the correspondence between topological edge modes and bulk invariants, we wish to use it in reverse: namely, by probing whether the bulk invariant is non-trivial, we wish to determine if the system hosts Majorana edge modes. To this end, we simulate the driven Kitaev chain at different driving frequencies. For each pair of frequencies $(\omega_1, \omega_2)$ and each Majorana mode ($\bar{\epsilon} = 0, \, \omega_1/2, \, \omega_2/2, \, (\omega_1 + \omega_2)/2$), we compute the localizer invariant, Eq.~(\ref{eq:invariant}), at the center of the chain, i.e. we use $x = N/2$, where $N$ is the length of the chain. If the local invariant is trivial ($C = 1$), then the corresponding Majorana mode is absent; if nontrivial ($C = -1$), then the system hosts Majoranas at the particular quasi-energy $\bar{\epsilon}$. Note that for each set of frequencies, we only need to compute the invariant at one location. Furthermore, since the matrix $M_x(\kappa)$ is in general sparse, there are readily available algorithms to efficiently compute the sign of its determinant (for example, using LU factorization).

Using the bulk invariant, we have compiled two phase diagrams, as shown in Fig.~\ref{fig:inv_freq}. Results from the spectral localizer approach indeed match our expectations, validating our real-space invariant for a time-quasiperiodic system. 

\textcolor{black}{In Fig.~\ref{fig:inv_freq}(a), we see that when $\omega_1$, $\omega_2$ exceed the system bandwidth $2J + \mu$, corresponding Majorana modes at $\bar{\epsilon} = \omega_1/2, \, \omega_2/2$ vanish. In contrast, the Majorana mode at zero quasi-energy always survives, regardless of the frequencies. Note that the red line in the lower-left square indicates that we are not considering cases where $\omega_1 \simeq \omega_2$. Along the direction perpendicular to $\boldsymbol{\omega} = (\omega_1, \omega_2)$, the energy variation across different synthetic lattice sites is very small.  Therefore, the quasi-periodic localization argument of Ref.~\cite{QPMajorana} is no longer valid, and the time-quasiperiodic Majoranas are not stable.}

\textcolor{black}{In Fig.~\ref{fig:inv_freq}(b), we further study the parameter regimes when the $(\omega_1 + \omega_2)/2$ Majorana mode is present by adding a third dynamic pairing potential $V((\omega_1 - \omega_2)t; \Delta_{12})$ to the Hamiltonian. Note that such a newly added term does not change the time quasiperiodicity of the model. We found that we are able to engineer the Majorana at energy $(\omega_1 + \omega_2)/2$ if $|\omega_1 - \omega_2|$ does not exceed $2J+\mu$, while both frequencies $\omega_{1,2}$ are comparable to $2J+\mu$, as expected. Similar to Fig.~\ref{fig:inv_freq}(a), we are not considering cases where $\omega_1 \sim \omega_2$, as indicated by the red line.}

\section{Conclusion and Outlook\label{sec:discussions}}
In this work, by exploiting a recently developed tool called spectral localizer,  we defined a real-space topological invariant at position $x$ as $C_{x,\bar{\epsilon}} = \text{sign}(\det(M_x(\kappa)))$, with non-Hermitian matrix $M_x(\kappa)= \kappa (X-xI) + i (K-\bar{\epsilon} I)$ and hyperparameter $\kappa$,  for time-quasiperiodic Majoranas at quasienergy $\bar{\epsilon}$ in a superconducting system driven at multiple incommensurate frequencies. Using the theory of PT-symmetric non-Hermitian matrices to analyze $M_x(\kappa)$, we provided detailed physical understanding of this invariant by analyzing the eigenvalues of $M_x(\kappa)$. We were able to interpret the maximal hyperparameter $\kappa_c$ beyond which our approach is invalid as the exceptional point for $M_x(\kappa)$.  Furthermore, we showed that despite the dense spectrum of frequency-space enlarged Hamiltonian $K$ for the time-quasiperiodic system, the invariat $C_{x,\bar{\epsilon}}$ only depends on th e real part of the eigenvalues of $M(\kappa)$, which only takes a finite number (linear in system size) of distinct values. We illustrated this approach numerically by studying a time-quasiperiodic driven Kitaev chain model.

In the future, we shall generalize this approach to other systems with dense spectrum, such as the gapless topological space-time crystals mentioned in Ref.~\cite{spacetime_crystal}. Another direction is to construct real-space invariant for interacting many-body systems, where the spectral localizer is not defined yet. However, our non-Hermitian matrix construction may provide some hints.
Finally, we notice that for small $\kappa$, the matrix $-iM_x(\kappa) = (K-\bar{\epsilon} I) - i\kappa (X - xI)$ is essentially a Hamiltonian with a small non-Hermitian perturbation,  which can be regarded as an effective Hamiltonian for the short-time dynamics in a open quantum system described by the Lindblad master equation~\cite{Ashida2020}. It would be interesting to explore the possibility of designing dissipative protocols for a driven Kitaev chain (or the Jordan-Wigner equivalent transverse-field Ising model) to measure the local topological invariant in experiments.

\begin{acknowledgments}
This work is supported by NSF PREP grant (PHY-2216774).
\end{acknowledgments}

\appendix
\section{Driven Kitaev Chain \label{app:kitaev}}
Under periodic boundary conditions, we may write the Hamiltonian Eq.~(\ref{eq:drivenchain}) with a two-frequency drive in momentum space as
\begin{equation}
    H = \sum_k \Psi_k^\dagger \left[ \mathcal{H}_K(k) + \mathcal{V}(k, \omega_1 t) + \mathcal{V}(k, \omega_2 t) \right] \Psi_k,
    \label{eq:kspace}
\end{equation}
where
\begin{gather}
   \mathcal{H}_K(k) =  \tau_z \xi_k + \tau_x \Delta \sin k \\
   \mathcal{V}(k, \omega t) = \tau_x \Delta' \sin k e^{i \omega t \sigma_z}.
\end{gather}
Here $\Psi_k^\dagger = (c_k^\dagger, c_{-k})$ is the Nambu spinor, $c_k = \sum_{j=1}^N c_j e^{-ikj}/\sqrt{N}$ is the annihilation operator in $k$ space, $\tau_{x,y,z}$ are Pauli matrices for Nambu space, and $\xi_k = -J \cos(k) - \mu/2$ is the normal state dispersion.

It has been shown in Ref.~\cite{QPMajorana} that the above system hosts four quasi-periodic Majoranas at quasi-energies $0$, $\omega_1/2$, and $\omega_2/2$.  Since quasi-energy is only defined modulo $n_1 \omega_1 + n_2 \omega_2$ in this system, under the particle-hole transformation, any state with one of the four quasi-energies is mapped to itself i.e. they are all particle-hole symmetric.

The robustness of quasi-periodic Majoranas may at first seem surprising. Indeed, since $\omega_1$ and $\omega_2$ are mutually irrational, $n_1 \omega_1 + n_2 \omega_2$ can approach any value for sufficiently large $|n_1|, |n_2|$, and the dense quasi-energy spectrum would naively imply mixing between the Majoranas and bulk states under local perturbation. Interestingly, despite the absence of a spectrum gap, quasi-periodic Majoranas are stable due to localization in the drive-induced synthetic dimensions. In fact, Ref.~\cite{QPMajorana} has shown that the Majoranas are robust even in the presence of temporal disorder and rational driving frequencies.

One may also understand the robustness of dynamical Majoranas from the system's quasi-energy band structure. In the absence of static and dynamic pairing potentials $\Delta = \Delta' = 0$, the quasi-energy bands of the system take the form $\epsilon_{n_1, n_2, e/h}(k) = \pm \xi_k + n_1 \omega_1 + n_2 \omega_2.$ For appropriate parameters, there can be at three special quasi-momenta $k_j$, $j \in \{0,1,2\}$, where the quasi-energy bands cross. Namely, the three special momenta satisfy
\begin{gather}
    \epsilon_{n_1, n_2, e}(k_0) = \epsilon_{n_1, n_2, h}(k_0) \nonumber \\ 
    \epsilon_{n_1, n_2, e}(k_1) = \epsilon_{n_1+1, n_2, h}(k_1) \nonumber \\
    \epsilon_{n_1, n_2, e}(k_2) = \epsilon_{n_1, n_2+1, h}(k_2). 
\label{eq:touching}
\end{gather}
Reinstating $\Delta, \Delta' \neq 0$ opens up topological gaps at these momenta. These gaps protect the Majorana modes at quasi-energies $0, \omega_1/2, \omega_2/2$, respectively.

Depending on the driving frequencies, the system can host different numbers of co-existing Majoranas and therefore be in different topological phases. The disappearance of Majorana modes at large frequencies is dictated by conditions in Eq.~(\ref{eq:touching}): when frequencies exceed the normal state bandwidth, there can no longer be quasi-energy band crossings. In turn, no topological gap opens up, and Majoranas lose their robustness and disappear. In terms of the system's parameters, Majorana modes at $\omega_1/2$, $\omega_2/2$ vanish when $\omega_1$, $\omega_2$ exceed the bandwidth $2J + \mu$, respectively. Note that the zero mode should always survive as long as $|\mu| < 2|J|$ in the presence of static pairing. 

\textcolor{black}{It is also possible to engineer a Majorana mode at energy 
\begin{equation}
(\omega_1 + \omega_2)/2 = (\omega_1 - \omega_2)/2 \mod \omega_2.
\end{equation}
One of the possibilities is to introduce a pairing term $\mathcal{V}(k;(\omega_1 - \omega_2)t)$ into the previous Hamiltonian. Following the same logic from the previous discussion, we need to make $|\omega_1 - \omega_2|$ smaller than the bandwidth $2J+\mu$. To make such a mode stable, we should also make $\omega_1,\omega_2 \gtrsim (2J+\mu)$, to make sure the Majorana at $(\omega_1 -\omega_2)/2$ does not overlap in energy with continuum states of the $\xi_k + n_1\omega_1 + n_2 \omega_2$ side bands.
}

\begin{figure}[ht]
    \centering
\includegraphics[width=0.8\linewidth]{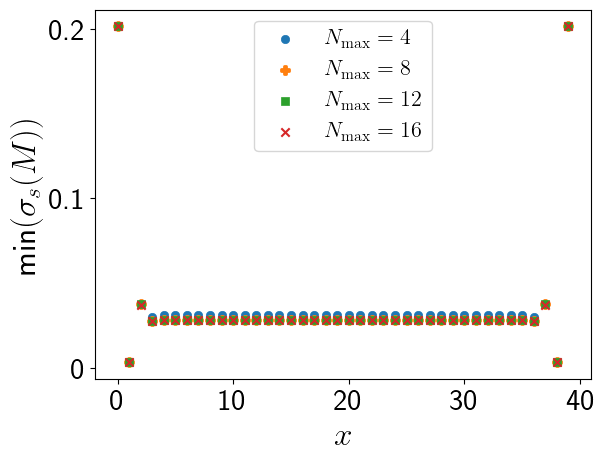}
    \caption{\textcolor{black}{The localizer gap  $\min(\sigma_s(M))$ along a chain for different number of frequency sectors $M$. The gap is independent of the cutoff and converges to a finite value. Parameter used are $J/\omega_1 = 0.5,\mu/\omega_1 = 1, \Delta/\omega_1 = 0.05, \Delta'/\omega_1 = 0.05, \kappa = 0.4, N = 40$, $\omega_2/\omega_1 = (\sqrt{5}+1)/2$. We have focused on the $\bar{\epsilon} = \omega_2/2$ mode.}}
    \label{fig:gapvsM}
\end{figure}
\begin{figure}[ht]
    \centering
    \includegraphics[width=0.8\linewidth]{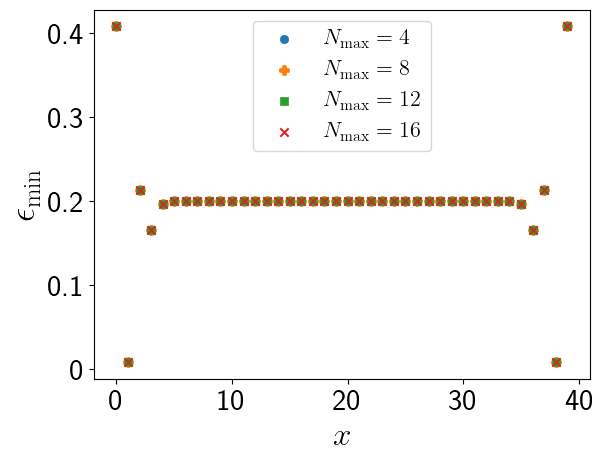}
    \caption{\textcolor{black}{Same parameters as Fig.~\ref{fig:gapvsM}, except for the quantity of interest is $\epsilon_{\text{min}}$, as defined in Eq.~(\ref{eq:epsilon_min}).}}
    \label{fig:epsilon_min}
\end{figure}

\section{Effects of Finite Frequency Truncation \label{app:MReal}}
In this Appendix, we show the robustness of the localizer gap and $\epsilon_{\text{min}}$ against truncation in the frequency domain.

For concreteness, we have again focused on the $\bar{\epsilon} = \omega_2/2$ mode. Therefore, when we truncate the frequency lattice to $N_{\text{max}}$ sectors, we restrict to $-N_{\text{max}} , \, ... \, , \, N_{\text{max}}$ harmonics of $\omega_1$ and $-N_{\text{max}}+1, \, ... \, , \, N_{\text{max}}$ harmonics of $\omega_2$ i.e. we are truncating the infinite frequency lattice to a $(2 N_{\text{max}} + 1) \times (2N_{\text{max}})$ lattice.


In Figs.~\ref{fig:gapvsM} and~\ref{fig:epsilon_min}, we show the localizer gap and  $\epsilon_{\text{min}}$ for different $N_{\text{max}}$. Note first that vanishing $\epsilon_{\text{min}}$ coincides with vanishing localizer gap (at $x=1$), as we had expected. In the limit of a dense spectrum $N_{\text{max}} \rightarrow \infty$, both $\epsilon_{\text{min}}$ and the localizer gap remain nonzero, indicating that a dense spectrum does not pose an issue, as discussed in Sec.~\ref{sec:dense}. Furthermore, both quantities converge easily for small cutoff $N_{\text{max}}$, showing our results only require a small frequency truncation.

\bibliography{main}

\end{document}